\pdfoutput=1
\documentclass[draft,grl]{agu2001}
\usepackage{graphicx}
\newcommand{\mbf}{\mathbf}
\newcommand{\vV}{\mbox{\boldmath $V$}}
\newcommand{\vE}{\mbox{\boldmath $E$}}
\newcommand{\vB}{\mbox{\boldmath $B$}}
\newcommand{\vj}{\mbox{\boldmath $j$}}
\authorrunninghead{JI ET AL.}

\titlerunninghead{Electron Dissipation in Reconnection}

\authoraddr{H. Ji,
Princeton Plasma Physics Laboratory, Princeton, NJ  08543, USA.
(hji@pppl.gov)}

\begin{document}

%
%
%
%

\title{New Insights into Dissipation in the Electron Layer During Magnetic Reconnection}

\authors{
H.~Ji, \altaffilmark{1,2}
Y.~Ren, \altaffilmark{1,3}
M.~Yamada, \altaffilmark{1,2}
S.~Dorfman,  \altaffilmark{1,2}
W.~Daughton, \altaffilmark{4}
and S.~P.~Gerhardt \altaffilmark{1,2}}

\altaffiltext{1}{Center for Magnetic Self-Organization in Laboratory and Astrophysical Plasmas}
\altaffiltext{2}{Princeton Plasma Physics Laboratory, Princeton, New Jersey, USA.}
\altaffiltext{3}{Department of Physics, University of Wisconsin - Madison, Madison, Wisconsin, USA.}
\altaffiltext{4}{Los Alamos National Laboratory, Los Alamos, New Mexico, USA.}

\begin{abstract}

Detailed comparisons are reported between laboratory observations of electron-scale dissipation layers near a reconnecting X-line and direct two-dimensional full-particle simulations.
Many experimental features of the electron layers, such as insensitivity to the ion mass, are reproduced
by the simulations;
the layer thickness, however, is about $3-5$ times larger than the predictions. Consequently, the leading candidate 2D mechanism based on collisionless electron nongyrotropic pressure is insufficient to explain the observed reconnection rates. These results suggest that, in addition to the residual collisions, 3D effects play an important role in electron-scale dissipation during fast reconnection.
\end{abstract}

%
%

%

\begin{article}

Despite the disruptive influences of magnetic reconnection on large-scale structures in plasmas,
the crucial topological changes and their associated dissipation 
take place only within thin current layers. 
The classical collisional models, 
where electrons and ions flow together through a single 
thin and long layer, fail to explain the observed fast reconnection rates.
Modern collisionless models predict~\citep{sonnerup79,mandt94,birn01} that 
ions exhaust through a thick, ion-scale layer while mobile electrons flow
through a thin, electron-scale layer,
allowing for efficient release of magnetic energy.
These ion layers have been frequently detected 
in space \citep[e.g.][]{deng01, oieroset01, mozer02} 
and studied in detail in the laboratory \citep{ren05,yamada06,brown06}.
In contrast, the electron layers, where magnetic field dissipates, are rarely encountered 
in space and are often detected at places far from the reconnection
X-line line~\citep{scudder02,mozer05,wygant05,phan07}.
Therefore, whether the electron layers indeed exist near the X-line, and if yes, 
whether their associated dissipation results predominantly from
laminar two-dimensional (2D) or three-dimensional (3D) dynamics
as suggested by \cite{xiao06,xiao07},
is still an open question.
Here we report detailed comparisons between recent laboratory observations 
of the electron layers near the X-line~\citep{ren08} and direct full-particle simulations in 2D.
The measured electron layers display properties strikingly similar to
predictions by 2D particle simulations,
including their geometrical shape, insensitivity to ion mass, 
and sensitivity to the boundary conditions, but disagree on
the electron layer thickness. As a consequence,
the leading 2D mechanism based on collisionless electron nongyrotropic pressure
is shown to be largely insufficient to explain the observed reconnection rates.
These results suggest that, in addition to the residual Coulomb collisions,
3D effects play an important role in electron-scale dissipation during fast reconnection.

The laboratory measurements were performed on the well
controlled and diagnosed experiment, Magnetic Reconnection
Experiment (MRX)~\citep{yamada97b}, as illustrated in Fig.1. 
A pair of coil assemblies, known as flux-cores, are
used to axisymmetrically initiate and maintain the reconnection process. 
Plasma is made by ionizing a pre-filled
gas through pulsing toroidal field coil current within the flux-cores 
during the period when the current flowing in the poloidal field (PF) 
coils peaks. When the PF coil current is ramped
down after the plasma is made, the field lines wrapped around both 
flux-cores are \lq\lq pulled" back, reconnect, and move towards the flux-cores.
Most of the important quantities can be either directly determined or
indirectly inferred from these measurements in cylinderical
coordinates ($R, Z, \theta$) assuming axisymmetry:  poloidal flux $\psi(R,Z,t)=\int_0^R 2\pi R^\prime
B_Z(R^\prime,Z,t)dR^\prime$ where $B_Z$ is the reconnecting
field; the toroidal reconnection electric field
$E_\theta=(\partial \psi/\partial t)/2\pi R$; 
and the toroidal current density $j_\theta \approx \mu_0^{-1} \partial B_Z/\partial R$. 
The density $n$ and electron temperature $T_e$ are
measured by a triple Langmuir probe and the flow speeds are
determined by a Mach probe. The typical plasma parameters
are: $n \simeq (0.1-2) \times 10^{20}$~m$^{-3}$, $T_e \sim T_i \simeq
(3-15)$~eV, $B<0.5$~kG.

Detection of the electron dissipation layer is made possible by taking
advantage of the differential motions between electrons and ions
or the so-called Hall effects~\citep{sonnerup79} in the reconnection region
without a guide field.  
These differential motions (or electric
current) within the reconnection plane produce out-of-plane
magnetic field component ($B_\theta$) with a quadrupole shape.
Conversely, accurate measurements of the $B_\theta$ profile
can determine the in-plane electron flow because of the much slower ion
flow in this region, and thus characterize the electron dissipation layer.
These measurements are performed
using five linear arrays of pickup coils (Fig.1); 
each array measures a one-dimensional profile of $B_\theta$ 
with a frequency response of 300kHz and with spatial resolutions up to 2.5 mm. 
This distance is close to the electron skin depth, $c/\omega_{pe}$ (=0.7-1.5mm)
where $\omega_{pe}$ is the electron plasma angular frequency,
and adequately resolves the electron layer whose minimum full thickness is 10 mm (see below).
These arrays are housed by thin glass tubes of outer diameter of 4 mm (four arrays) or 
5 mm (one array) with shielding from electrostatic noise. 
The presence of these probes in the plasma does not appear to affect the reconnection process, but it may cause modest overestimates of the electron layer thickness (see below).

One such example measurement is shown in Fig.2(b)
where the in-plane electron flow ($V_{eZ}$ and $V_{eR}$) is shown as arrows
while the normalized, out-of-plane magnetic field is shown
as color-coded contours in the left half of the reconnection plane. 
Electron outflow speed, $V_{eZ}$, is also shown as functions of $Z$ in Fig.2(c)
(at the current sheet center) and $R$ in Fig.2(a) (across the reconnection region 
at the location where $V_{eZ}$ peaks).
The dimensions of the electron layer
can be characterized by the half thickness $\delta_e$ (the radial distance 
during which $V_{eZ}$ decreases by 60\% from its peak value) and the half length $L_e$ 
(the axial distance during which $V_{eZ}$ increases from zero to its peak).
We positively identify this region as the electron dissipation layer 
because both its dimensions, $\delta_e$ and $L_e$, are independent
of ion mass, as shown in Fig.3 for $\delta_e$. 

Dissipation in the electron layer is governed by the electron equation of motion,
\begin{equation} 
m_e n \left( {\partial \over \partial t} + \vV_e \cdot \nabla \right)  \vV_e
 =  -en \left(\vE + \vV_e \times \vB \right) -\nabla \cdot {\mbf P}_e 
+ en\eta_{Spitzer} \vj ,
\end{equation}
where $m_e$ is electron mass, ${\mbf P}_e$ electron pressure tensor,
and $\eta_{Spitzer}$ the Spitzer resistivity due to Coulomb collisions with ions~\citep{spitzer62}.
In the modern collisionless steady-state 2D models,
the reconnection electric field, $E_\theta$, can be only possibly
balanced by either the Hall term $(\vV_e \times \vB)_\theta \approx (\vj \times \vB)_\theta /en$, 
the inertia terms, 
or the electron pressure tensor term $(\nabla \cdot {\mbf P}_e)_\theta$.
While the Hall term is important in supporting $E_\theta$ within the ion layer~\citep{birn01,ren05,yamada06},
it diminishes within the electron layer especially near the X-line due to the vanishing $\vB$.
It has been shown in particle simulations~\citep{cai97,hesse99,pritchett01,kuznetsova01} 
that $E_\theta$ is supported primarily by the electron pressure tensor term
following earlier suggestions~\citep{vasyliunas75}.
This mechanism has been since widely accepted as the leading candidate
to provide the required dissipation within the electron layer. 
It is, however, extremely difficult to confirm
this pressure anisotropy, directly or indirectly, by measurements in real plasmas~\citep{scudder02}.

One of the predictions of these 2D particle simulations 
is that the half thickness of the electron layer, $\delta_e$, scales
as $(1-2)c/\omega_{pe}$~\citep{pritchett01}. 
The measured $\delta_e$ in MRX, however, scales
as $\sim 8c/\omega_{pe}$ (Fig.3).
Current blockage due to the probes is estimated to lead to a $6-44\%$ 
increase in the measured $\delta_e$, depending on the ratio of $\delta_e$ 
to the glass tube radius.
Applying these corrections leads to $\delta_e = (5.5-7.5) c/\omega_{pe}$.
To better compare with the experiment, on the other hand,
we have constructed a kinetic numerical model~\citep{dorfman08} 
using boundary conditions similar to the MRX based on the existing NPIC 2D code~\citep{daughton06}.
A 75cm$\times$150cm simulation box 
is used with conducting boundary conditions for fields and elastic reflection for particles at the walls. 
Two current carrying coils of radius 1.3 cm are contained 
within a larger concentric flux core of radius 9.4 cm.  The flux cores are spaced 40 cm apart as in the experiment. 
The flux core surface is approximated as an insulating boundary; 
particles may be absorbed or reflected. 
Due to constraints on computation resources, the number of the Debye lengths per 
$c/\omega_{pe}$
is limited compared to the experiment, but there is strong evidence that 
the reconnection rate and electron layer scalings are insensitive to this number as long as the initial plasma beta is fixed~\citep{dorfman08}.
As the current is ramped down according to a sinusoidal waveform modeled on the PF coil current of MRX and reconnection is driven,
both ion and electron dissipation layers are formed.
Simulation parameters are chosen such that the global
reconnection rate and the current sheet thickness on the ion scale match the observations. 
An example run is shown in Fig.2(d-f) in the same format 
as in Fig.2(a-c), and most of the observed features, including
geometrical shapes and out-of-plane magnetic component, are reproduced.

The quantitative agreement between experiment and simulation is, however, found for only 
the global ion dynamics but not the local electron dynamics. 
The experimentally observed independence of $\delta_e$ and $L_e$ 
on ion mass was reproduced as shown for $\delta_e$ by the open squares in Fig.3
for a fixed but artificially heavy electron mass.
The values of $\delta_e$ in units of $c/w_{pe}$ (evaluated using a line-averaged density at $Z=0$), 
however, are much smaller in simulations than in experiments, as illustrated by an
alternative ordinate in Fig.2(a) and (d). In Fig.3,
a case at higher mass ratio (400) with a different electron mass is also plotted along with simulations with a realistic hydrogen mass ratio but a smaller simulation domain and open boundary conditions~\citep{daughton06}.  All of these cases, including more recent simulations using different open boundary conditions~\citep{huang08}, confirm a linear relation of $\delta_e =(1.5-2)c/\omega_{pe}$ which is about $3-5$ times thinner than the experiment.  In contrast, the dependence of the length of the electron layer ($L_e$) on $c/\omega_{pe}$ is less robust; it can change significantly when the reflection coefficient parameter on the flux core surface is varied~\citep{dorfman08} as expected from the observed dependence of the reconnection process on boundary conditions~\citep{kuritsyn07}.

The fact that the observed electron layers are substantially thicker than the numerical predictions 
implies different dissipation mechanisms operating between these two cases.
In fact, our collisionless simulation model
does not include the residual collisions between electrons and ions or neutrals. 
But in MRX only a fraction of $E_\theta$ can be accounted
for by the classical resistivity, $E_\eta \equiv \eta_{Spitzer} j_\theta$ (Fig.4).
Collisions between electrons and neutrals, and electron collisional viscous effects
are also estimated to be unimportant in these discharges with low fill pressure.
The electron inertia terms, $(m_e/e)[(V_{eR} \partial /\partial R)+(V_{eZ} \partial /\partial Z)]V_{e\theta}$,
are estimated to be on the order of 1 V/m, which is negligibly small.
Near the X-line, the effects due to electron nongyrotropic pressure can be well approximated by~\citep{hesse99}
\begin{equation}
E_{NG} \equiv -\left( \nabla \cdot {\mbf P}_e \over en\right)_\theta
\approx{1\over e} {\partial V_{eZ} \over \partial Z} \sqrt{2 m_e T_e},
\end{equation}
as also validated in our kinetic model. Direct evaluations of $E_{NG}$ using the
measured profile, $V_{eZ}(Z)$ as in Fig.2, 
gives values only a small fraction of $E_\theta-E_\eta$ (Fig.4). 
This leaves the majority of $E_\theta$ still unexplained, and therefore 
there must exist additional dominant dissipation mechanisms. 
Because our kinetic model contains all possible collisionless kinetic mechanisms operative in 2D, 
these dominant mechanisms must be 3D in character,
including effects due to current sheet deformation or plasma turbulence through wave-particle interactions within the current sheet.
The latter was indeed already suggested by the detection of electromagnetic fluctuations~\citep{ji04}
when dissipation increases at low collisionalities~\citep{ji98}.
This subject is also under intensive theoretical and numerical investigation,
such as recently by \cite{moritaka07},
in the search for mechanisms for fast reconnection. Lastly, we comment that 
these 3D effects, in additional to the residual collisions, may diffuse substantially
the predicted two-scale structures seen in the profiles of the reconnecting magnetic field, 
which remain undetected thus far in the experiment.

\begin{acknowledgments}
The authors thank R. Kulsrud for the insightful discussions. 
The MRX project is supported by DOE, and the numerical comparisons
reported here are mainly supported by the NASA Geosciences Program.
SD was supported by the Fusion Energy Sciences Fellowship Program
and the NDSEG program.
\end{acknowledgments}

\begin{figure}
\centerline{\includegraphics[width=2.5truein]{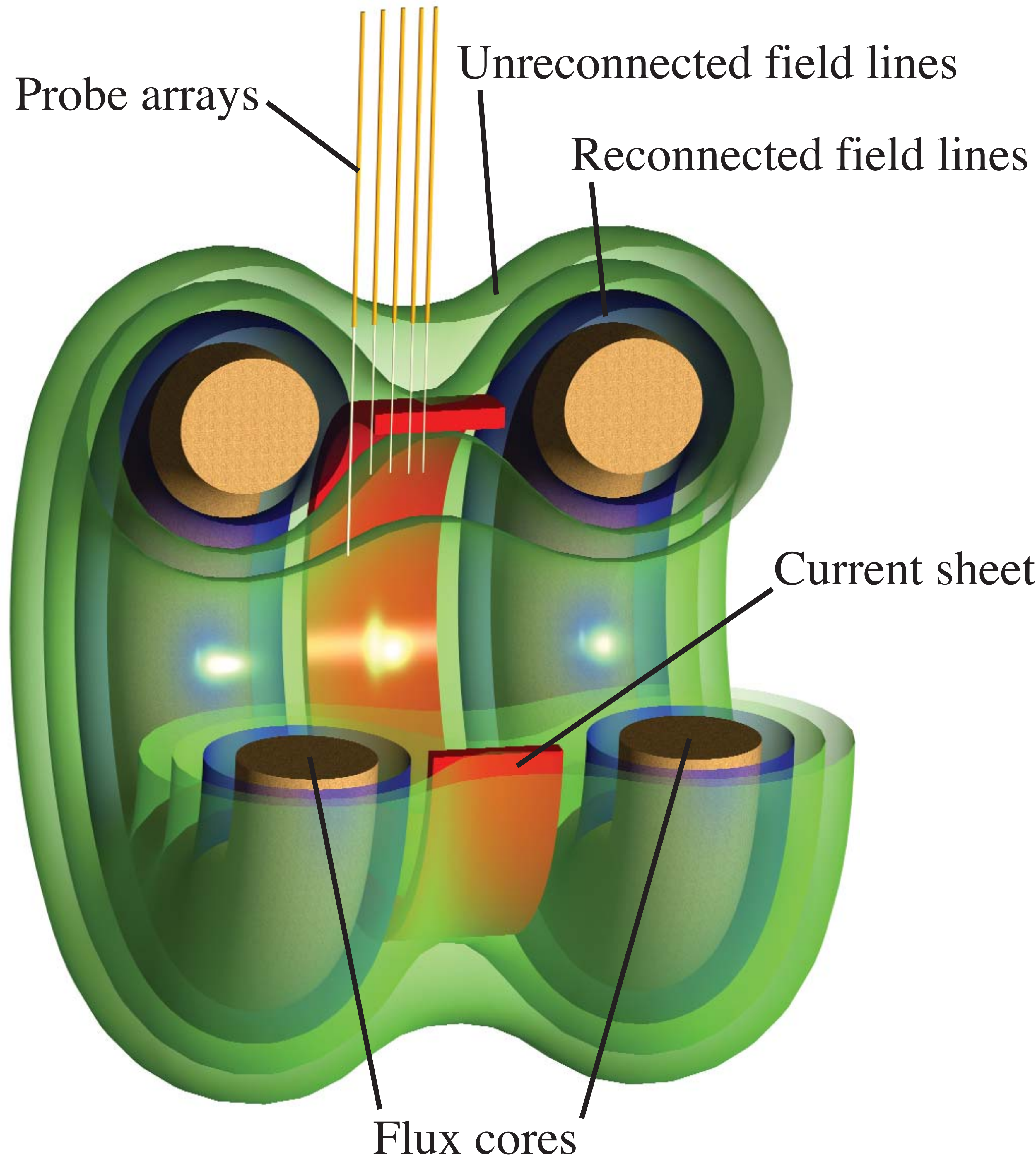}}
\caption{Experimental set-up of MRX device. The toroidal direction points along the current sheet
while the poloidal direction wraps around the flux cores.}
\end{figure}

\begin{figure}
\centerline{\includegraphics[scale=0.45]{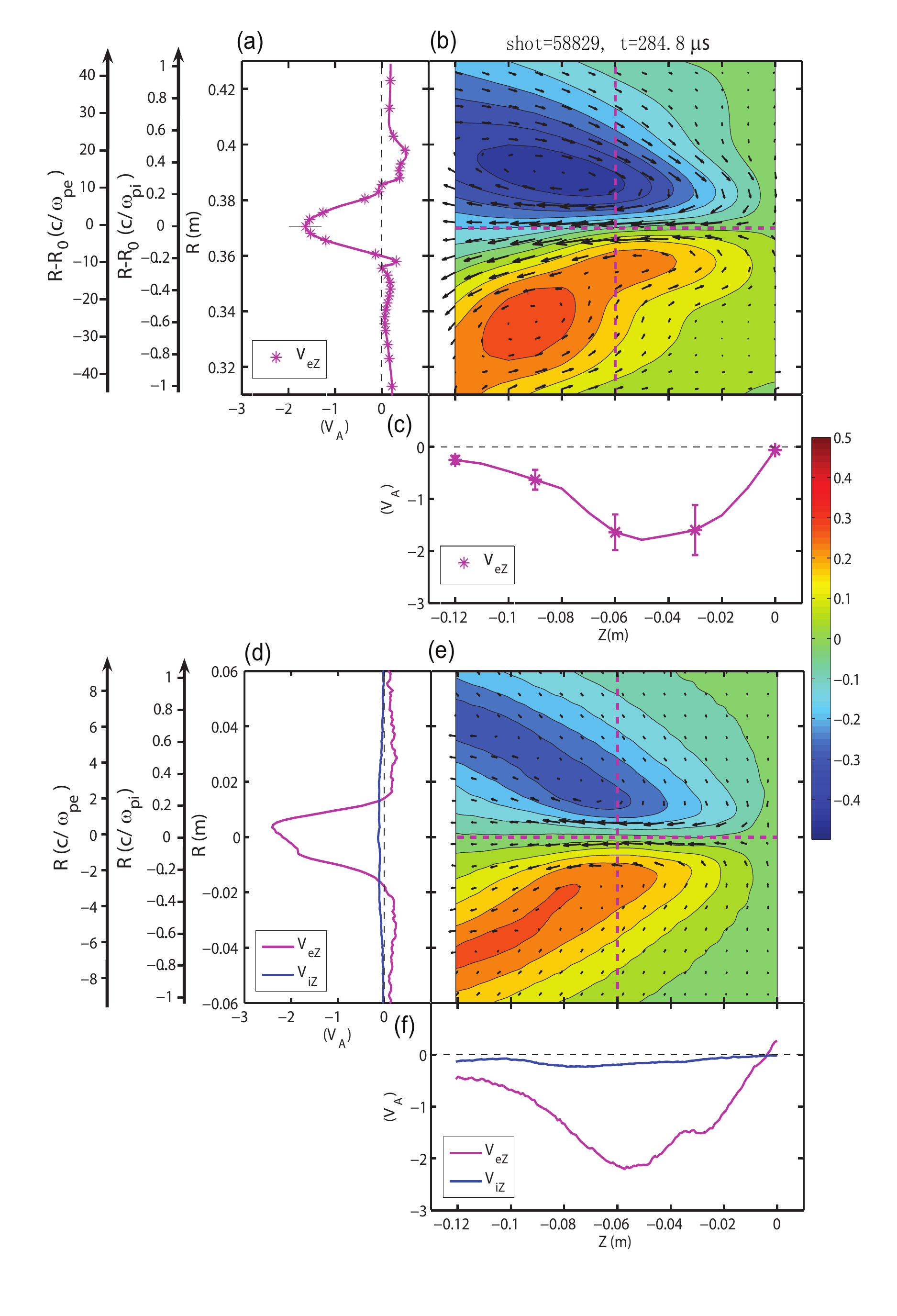}}
\caption{Identification of electron dissipation layer.
The top three panels (a-c) show an experimental example
taken from a hydrogen plasma with a fill pressure of 2 mTorr.
Results from a corresponding simulation are shown in the same format 
in the lower three panels (d-f).
The parameters used in the simulation are: 
$864 \times 1728$ cells with 0.5 billion particles per species,
initial density of $2.6\times 10^{19} {\rm m}^{-3}$, $m_i=m_{\rm hydrogen}$,
$m_e=m_{\rm hydrogen}/75$, a time scale for the coil current ramp down is
185 initial ion cyclotron times, and no particle reflections at the flux core surface.}
\end{figure}

\begin{figure}
\centerline{\includegraphics[scale=0.27]{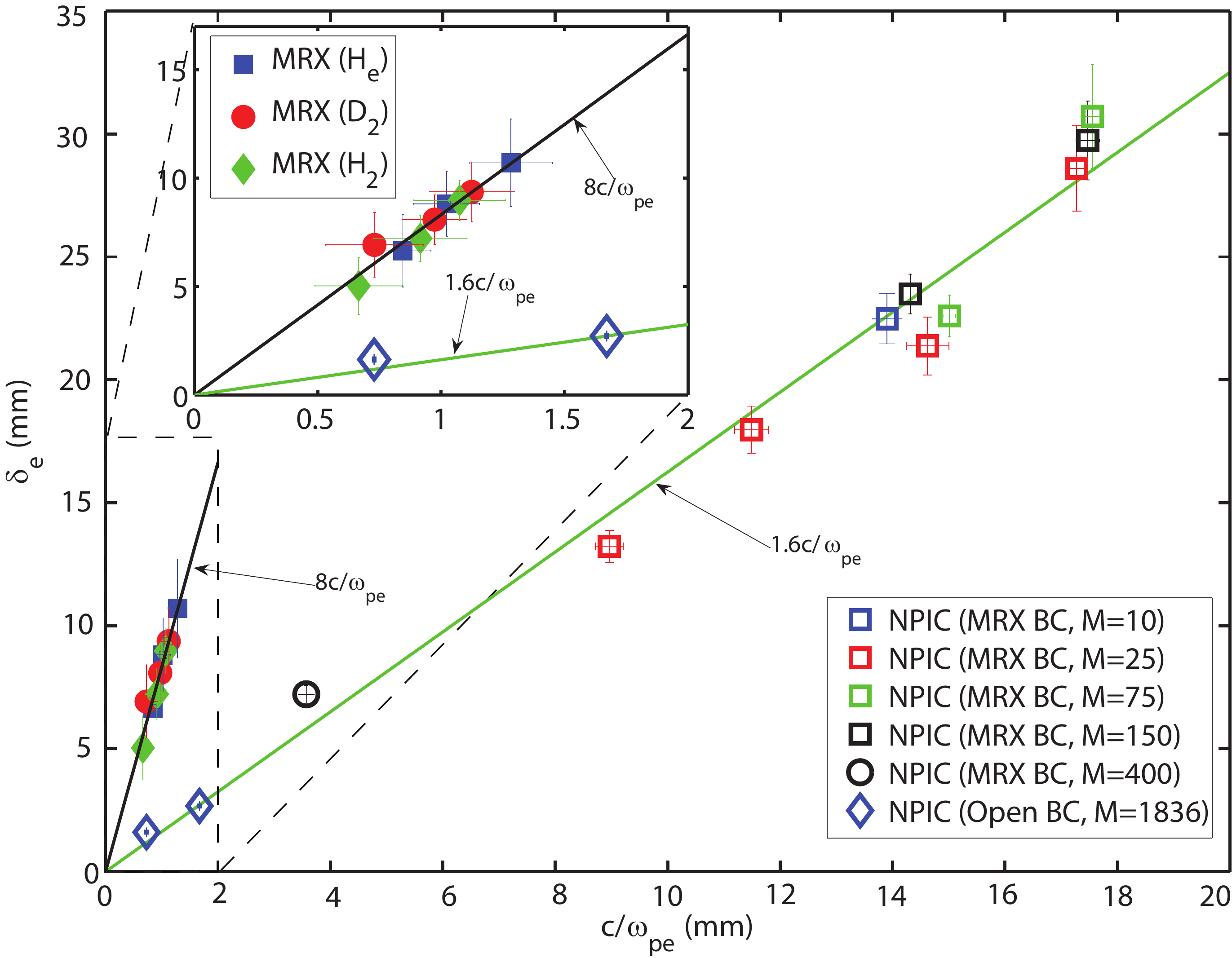}}
\caption{Scaling of width of electron dissipation layer. Filled symbols show
the experimentally measured $\delta_e$ as a function of the
electron skin depth ($c/\omega_{pe}$) for three different ion
species. The error bars result mainly from shot-to-shot variations.
Open symbols show $\delta_e$ determined from 2D PIC simulations.}
\end{figure}

\begin{figure}
\centerline{\includegraphics[scale=0.35]{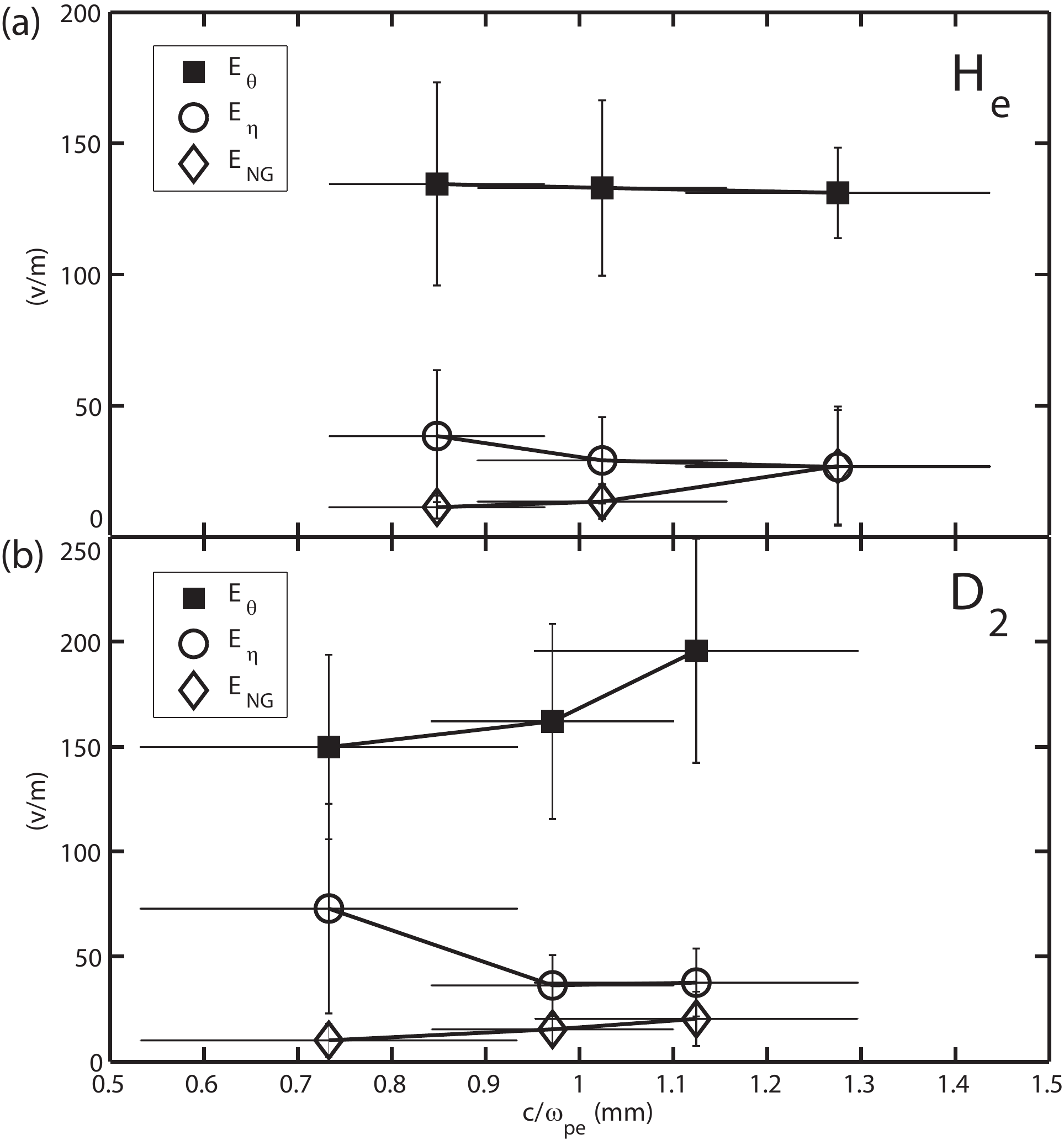}}
\caption{Composition of reconnecting electron field, $E_\theta$, 
for (a) helium and (b) deuterium plasmas. 
Total reconnecting electric field in MRX, $E_\theta$, and the part of it
due to electron-ion collisions, $E_\eta=\eta_{Spitzer} j_\theta$ near the X-line are plotted
as a function of $c/\omega_{pe}$. The estimated electric field due to
electron nongyrotropic pressure, $E_{NG}$, is also shown.}
\end{figure}

%
%
%
%
%
%
%
%



%
%
%
\end{article}

\end{document}